\documentclass[conference,a4paper]{IEEEtran}

\usepackage[utf8]{inputenc} 
\usepackage[T1]{fontenc}
\usepackage{url}
\usepackage{hyperref}
\usepackage{ifthen}
\usepackage{cite}
\usepackage{stfloats}
\usepackage{lipsum}
\usepackage{cuted}
\usepackage[cmex10]{amsmath} % Use the [cmex10] option to ensure complicance
\usepackage[pdftex]{graphicx}
\ifCLASSOPTIONcompsoc
  \usepackage[caption=false,font=normalsize,labelfont=sf,textfont=sf]{subfig}
\else
\usepackage[caption=false,font=footnotesize]{subfig}
\fi
\usepackage{enumitem}
\DeclareGraphicsExtensions{.pdf}
\usepackage{microtype}
\usepackage{mathtools}

\usepackage{amssymb}
\newtheorem{theorem}{Theorem}

\newtheorem{lemma}{Lemma}
\newtheorem{remark}{Remark}

\begin{document}
\title{Can Marton Coding Alone Ensure Individual Secrecy?} 

% %%% Single author, or several authors with same affiliation:
\author{%
   \IEEEauthorblockN{Jin Yeong Tan, Lawrence Ong, and Behzad Asadi}
   \IEEEauthorblockA{School of Electrical Engineering and Computing, The University of Newcastle, Newcastle, Australia\\
                     Email: jinyeong.tan@uon.edu.au, lawrence.ong@newcastle.edu.au, behzad.asadi@uon.edu.au}
}

\maketitle

%%%%%%
%% Abstract: 
%% If your paper is eligible for the student paper award, please add
%% the comment "THIS PAPER IS ELIGIBLE FOR THE STUDENT PAPER
%% AWARD." as a first line in the abstract. 
%% For the final version of the accepted paper, please do not forget
%% to remove this comment!
%%

\begin{abstract}
	For communications in the presence of eavesdroppers, random components are often used in code design to camouflage information from eavesdroppers. In broadcast channels without eavesdroppers, Marton error-correcting coding comprises random components which allow correlation between auxiliary random variables representing independent messages. In this paper, we study if Marton coding alone can ensure individual secrecy in the two-receiver discrete memoryless broadcast channel with a passive eavesdropper. Our results show that in accordance to the principle of Wyner secrecy coding, this is possible and Marton coding alone guarantees individual secrecy. However, this comes with a penalty of requiring stricter channel conditions.
\end{abstract}


%% The paper must be self-contained. However, if you are referring to
%% a full version for checking certain proofs, please provide the
%% publically accessible location below.  If the paper is completely
%% self-contained, you can remove the following line from your
%% submission.

\section{Introduction}\label{S1}

\subsection{Background}\label{S1A}

The problem of secure communication over broadcast channels is always of great importance since broadcast channels are widely applicable in wireless communication systems. Some popular works on secure broadcast channel have been presented by Csisz$\acute{\text{a}}$r and K$\ddot{\text{o}}$rner~\cite{iref1}, Chia and El Gamal~\cite{iref2} as well as Schaefer and Boche~\cite{Schaefer_Boche14}. These works~\cite{iref1,iref2,Schaefer_Boche14} studied cases of two- or three-receiver broadcast channels in which a common message is transmitted to all receivers and a private message is protected from certain receivers. Wyner secrecy coding~\cite{cref4} is employed to achieve security therein. The complexity of the problem increases as we consider the protection of two private messages from an eavesdropper which is not one of the receivers. In this case, Chen et al.~\cite{cref1} proposed a secrecy coding scheme which combines Wyner secrecy coding~\cite{cref4} and Carleial-Hellman secrecy coding~\cite{cref3}; whereas Mansour et al.~\cite{Mansour_Schaefer_Boche16} proposed a secrecy coding scheme which combines Wyner secrecy coding~\cite{cref4} and one-time pad~\cite{cref8}. The works discussed above were also extended to more specific broadcast channels with channel states. For instance, several studies looked into secure communications in broadcast channels with receiver side information at the legitimate receivers which is unknown to the eavesdropper~\cite{cref2,iref13,iref14,Me1,Me2}. 

From the works discussed above, one can easily notice that regardless of the channel setup, it seems to be a norm for secure broadcasting to be achieved by integrating secrecy techniques into error-correcting coding schemes. Among the secrecy techniques covered in these works are the secret key approach (also known as one-time pad)~\cite{cref8}, Wyner secrecy coding~\cite{cref4} and Carleial-Hellman secrecy coding~\cite{cref3}. When secure broadcasting is necessary, these secrecy techniques are often integrated independently or as a combination to common error-correcting coding scheme for broadcast channels such as the superposition coding scheme~\cite{cref6} and Marton coding scheme~\cite{cref5}. As a consequence of this common practice, there is little knowledge on whether secure broadcasting can be achieved by having only an error-correcting coding scheme. 

In regards to this, we have only come across some brief insights in two works. Although two distinct secrecy coding schemes have been proposed by Chen et al.~\cite{cref1} and Mansour et al.~\cite{Mansour_Schaefer_Boche16} for the two-receiver discrete memoryless broadcast channel with a passive eavesdropper, both schemes presented the usage of randomness in error-correcting coding schemes to ensure secrecy. In particular, the random components in Marton coding were utilized to help in message protection. However, this protection can only be achieved when the random components in Marton coding are complemented with other secrecy techniques.

This motivates us to explore if the dependence on additional secrecy techniques in secrecy coding scheme construction can be removed, which means it is possible to use the randomness in Marton coding alone to provide secure communication. It is interesting to see how the Marton coding scheme alone extends itself into the area of secure broadcasting and leads us to new ways of constructing secrecy coding schemes. Throughout this paper, we will also be considering the individual secrecy notion which requires the individual information leakage from each message to the eavesdropper to be vanishing~\cite{iref14,iref15,cref1,cref2}. In short, this paper aims to study if Marton coding alone can ensure individual secrecy and derive the corresponding individual secrecy rate region for the two-receiver discrete memoryless broadcast channel with a passive eavesdropper. 

\subsection{Contributions}\label{S1B}

Since the usage of only error-correcting coding schemes to ensure secure communication has yet been attempted across any literature in our knowledge, it is necessary for us to identify a proper starting point to our work. We notice that having random components, the Marton coding scheme~\cite{cref5} appears to share structural similarities with the Wyner secrecy coding scheme~\cite{cref4}. Since Wyner secrecy coding is a popular secrecy coding technique that has been widely applied to provide information protection, we draw the hypothesis that Marton coding alone should be able to guarantee secrecy as well. In this paper, we prove that the Marton coding scheme alone can provide message protection under the notion of individual secrecy. This is possible when certain channel constraints are satisfied during codebook generation. Using this strategy, we derive an inner bound for the two-receiver discrete memoryless broadcast channel with a passive eavesdropper. The ability of Marton coding in ensuring secure communication without the need of additional secrecy techniques is a great observation since it may lead to the construction of simpler and more effective secrecy coding schemes in future works.   
%We also obtain an answer to the contradicting ideas raised by Chen, Koyluoglu and Sezgin~\cite{cref1} and Mansour, Schaefer and Boche~\cite{Mansour_Schaefer_Boche16}. 

\subsection{Paper Organization}\label{S1C}

The entire paper will be organized as follows. Section II will focus on the system model. Section III will provide the main results on using Marton coding to ensure individual secrecy. Next, Section IV will present some discussions and conclude the paper. 

\section{System Model}\label{S2}

In this paper, we will denote random variables by uppercase letters, their corresponding realizations by lowercase letters and their corresponding sets by calligraphic letters. 
%A $(j-i+1)$-sequence of random variables will be denoted by $X_i^j=(X_i,…,X_j)$ for $1\leq i\leq j$. Whenever $i=1$, the subscript will be dropped, resulting in $X^j=(X_1,…,X_j)$.%
A $n$-sequence of random variables will be denoted by $X^n=(X_1,…,X_n)$, where $X_i$ represents the $i$th variable in the sequence. $\mathbb{R}^d$ represents the $d$-dimensional real Euclidean space and $\mathbb{R}_+^d$ represents the $d$-dimensional non-negative real Euclidean space. $\mathcal{R}$ will be used to represent a subset of $\mathbb{R}^d$. $\mathbb{Z}$ represents the set of all integers. $\mathcal{T}^{(n)}_{\epsilon}$ represents the set of jointly $\epsilon$-typical $n$-sequences. Meanwhile, $[a:b]$ refers to a set of natural numbers between and including $a$ and $b$, for $a\leq b$. Lastly, the operator $\times$ denotes the Cartesian product.

\begin{figure}[t]
	\centering
	\includegraphics[scale=0.55]{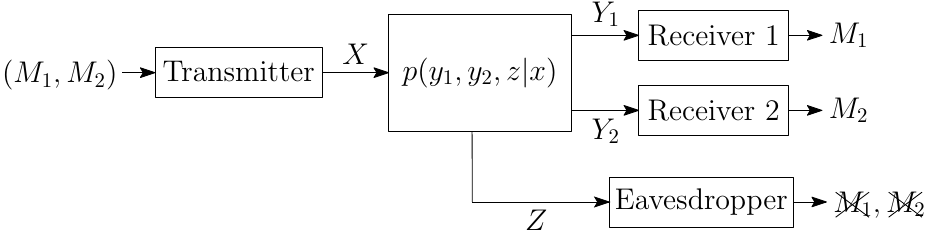}
	\caption{The two-receiver broadcast channel in the presence of an eavesdropper.}
	\label{fig:1}
\end{figure}

The paper focuses on the two-receiver discrete memoryless broadcast channel with a passive eavesdropper. The system model for this case is illustrated in Fig.~\ref{fig:1}. In this model, we define $(M_1, M_2)$ as the source messages, $M_i$ as the message requested by legitimate receiver $i$, for all $i\in\{1,2\}$. Let $X$ denote the channel input from the sender, while $Y_i$ and $Z$ denote the channel output to receiver $i$ and the eavesdropper respectively. In $n$ channel uses,  $X^n$ represents the transmitted codeword, $Y_i^n$ represents the signal received by legitimate receiver $i$ and $Z^n$ represents the signal received by the eavesdropper. The memoryless (and without feedback) nature of the channel also implies that
\begin{align}
	&p(y_1^n,y_2^n,z^n|x^n)=\prod_{i=1}^n p(y_{1i},y_{2i},z_i|x_i).
\end{align} 

In this case, the transmitter will be sending messages $M_{1}$ and $M_{2}$ to legitimate receiver 1 and 2, respectively through the channel $p(y_{1},y_{2},z|x)$. Besides, both messages $M_{1}$ and $M_{2}$ need to be individually protected from the eavesdropper.

\textit{Definition 1:} A $(2^{nR_1},2^{nR_2},n)$ secrecy code for the two-receiver discrete memoryless broadcast channel consists of:
\begin{itemize}
	\item two message sets, where $\mathcal{M}_1=[1:2^{nR_1}]$ and $\mathcal{M}_2=[1:2^{nR_2}]$;
	\item an encoding function, $f:\mathcal{M}_1\times\mathcal{M}_2\rightarrow\mathcal{X}^n$, such that $X^n=f(M_1,M_2)$; and
	\item two decoding functions, where $g_1:\mathcal{Y}_1^n\rightarrow\mathcal{M}_1$, such that $\hat{M}_1=g_1(Y_1^n)$ at receiver 1 and $g_2:\mathcal{Y}_2^n\rightarrow\mathcal{M}_2$, such that $\hat{M}_2=g_2(Y_2^n)$ at receiver 2.
\end{itemize}

Both messages, $M_1$ and $M_2$ are assumed to be uniformly distributed over their respective message set. Hence, we have $R_i=\frac{1}{n}H(M_i)$, for all $i\in\{1,2\}$. Meanwhile the individual information leakage rate associated with the $(2^{nR_1},2^{nR_2},n)$ secrecy code is defined as $R_{\text{L},i}^{(n)}=\frac{1}{n}I(M_i;Z^n)$, for all $i\in\{1,2\}$. The probability of decoding error for the secrecy code at each receiver $i$ is defined as $P_{\text{e},i}^{(n)}=\mathrm{P}\{\hat{M}_i\ne M_i\}$, for $i\in\{1,2\}$. 

A rate pair $(R_1,R_2)$ is said to be achievable if given any $\epsilon>0$ and $\tau >0$, there exists $n'$ such that for all $n>n'$, there exists a sequence of $(2^{nR_1},2^{nR_2},n)$ codes satisfying 
\begin{align}
	&P_{\text{e},i}^{(n)}\leq \epsilon\text{, for all }i\in\{1,2\} \text{ and }\label{dec}\\			
	&R_{\text{L},i}^{(n)}\leq \tau\text{, for all }i\in\{1,2\}.\label{sec}		
\end{align}

\section{Bridging Marton Coding and Wyner Secrecy Coding}\label{S3}

\begin{figure}[t]
	\centerline{
		\subfloat[Marton coding scheme]{\includegraphics[scale=0.5]{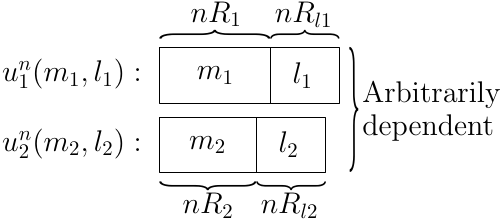}
		% where an .eps filename suffix will be assumed under latex,
		% and a .pdf suffix will be assumed for pdflatex
		\label{fig:2a}}}
	\centerline{
		%\hfil
		\subfloat[Wyner secrecy coding scheme]{\makebox[7cm][c]{\includegraphics[scale=0.5]{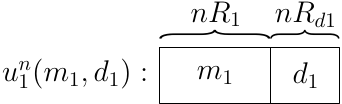}}
		% where an .eps filename suffix will be assumed under latex,
		% and a .pdf suffix will be assumed for pdflatex
		\label{fig:2b}}}
	\caption{Comparison between Marton coding scheme and Wyner secrecy coding scheme.}
	\label{fig:2}
\end{figure}

In this section, we will establish a connection between Marton coding~\cite{cref5} and Wyner secrecy coding~\cite{cref4}. More precisely, we will show that Marton coding alone can provide message protection under certain channel conditions even without additional secrecy techniques. 

\subsection{Brief Review of Marton Coding and Wyner Secrecy Coding}\label{S3A}

Prior to our actual discussion, we will review the Marton coding scheme~\cite{cref5} and the Wyner secrecy coding scheme~\cite{cref4} which are illustrated in Fig.~\ref{fig:2}. The Marton coding scheme is used while transmitting messages to multiple receivers through the broadcast channel. Considering the two-receiver broadcast channel whereby the transmitter sends $M_1\in [1:2^{nR_{1}}]$ and $M_2\in [1:2^{nR_{2}}]$ to receiver 1 and 2 respectively, we see from Fig.~\ref{fig:2a} that the Marton coding scheme comprises two codeword layers $U_1^n$ and $U_2^n$ which carry $M_1$ and $M_2$ respectively. The codeword layers also contain the random components $L_1\in [1:2^{nR_{l1}}]$ and $L_2\in [1:2^{nR_{l2}}]$ which can be chosen to allow $U_1^n$ and $U_2^n$ to be arbitrarily dependent while $M_1$ and $M_2$ are independent.

Meanwhile, the Wyner secrecy coding scheme is used to transmit a message to a single receiver through the wiretap channel in which the message needs to be kept protected from an eavesdropper. As in Fig.~\ref{fig:2b}, the Wyner secrecy coding scheme comprises the codeword layer $U_1^n$ which carry $M_1$. The Wyner random component $D_1\in [1:2^{nR_{d1}}]$ is required to ensure that the eavesdropper gains almost no information about the message sent~\cite{cref4}. 

\subsection{An Intuition and the Challenges}\label{S3B}

Our idea to establish a connection between Marton coding and Wyner secrecy coding originates from a simple observation that both schemes share structural similarities. 
Notice that both schemes transmit the desired message to the receivers via their respective codewords. Besides, the random components in Marton coding $L_1$ and $L_2$ also share similarties with the Wyner random component $D_1$ since they are formed by generating additional sequences for each message. These similarities thus beg the question: If the Wyner secrecy coding scheme is capable of ensuring secrecy, shouldn't the same apply to the Marton coding scheme?

\begin{figure}[t]
	\centerline{
		\subfloat[Codebook of the Marton coding scheme]{\makebox[7cm][c]{\includegraphics[scale=0.6]{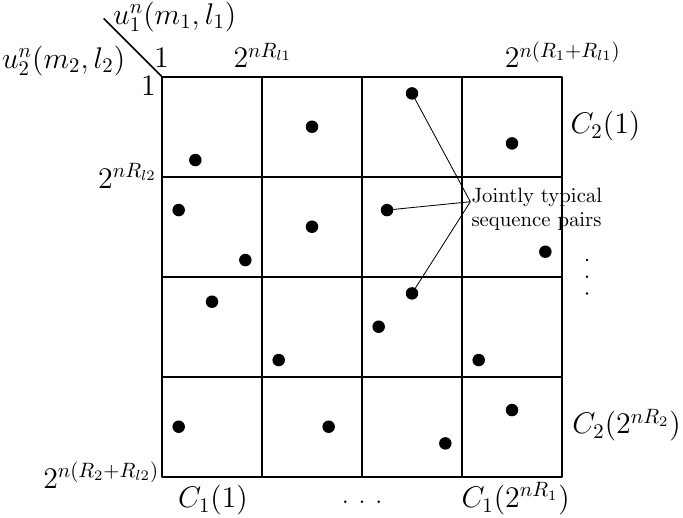}}
		% where an .eps filename suffix will be assumed under latex,
		% and a .pdf suffix will be assumed for pdflatex
		\label{fig:6a}}}
	\centerline{
		%\hfil
		\subfloat[Codebook of the Wyner secrecy coding scheme]{\makebox[7cm][c]{\includegraphics[scale=0.6]{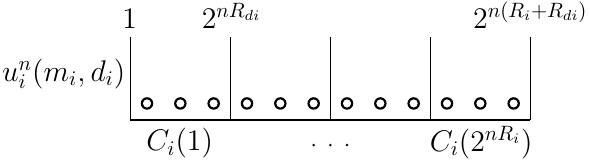}}
		% where an .eps filename suffix will be assumed under latex,
		% and a .pdf suffix will be assumed for pdflatex
		\label{fig:6b}}}
	\caption{Codebook comparison between the Marton coding scheme and the Wyner secrecy coding scheme.}
	\label{fig:6}
\end{figure}

The answer to this question is not as straightforward as it seems. Although both schemes have similar structures, there is a slight difference during codebook generation which consequently affects the encoding. Applying Wyner secrecy coding to the broadcast channel~\cite{iref1}, as seen in Fig.~\ref{fig:6b}, we will be generating a subcodebook $C_i(m_i)$ for each message $m_i\in [1:2^{nR_{i}}]$, $i=1,2$. Each subcodebook consists of $2^{nR_{di}}$ randomly and independently generated sequences $u_i^n(m_i,d_i)$, $d_i\in [1:2^{nR_{di}}]$. This codebook will then be revealed to all parties.

On the other hand, as seen in Fig.~\ref{fig:6a}, the Marton coding scheme~\cite{cref5} undergoes similar codebook generation steps, resulting in subcodebooks which each consists of $2^{nR_{li}}$ randomly and independently generated sequences $u_i^n(m_i,l_i)$, $l_i\in [1:2^{nR_{li}}]$. The difference sets in when the Marton coding scheme requires the $U_1^n$ and $U_2^n$ codeword layers to be dependent according to some chosen joint distribution $p_{U_1U_2}$. This requires a sequence pair $(u_1^n(m_1,l_1),u_2^n(m_2,l_2))$ to be preselected in each product subcodebook $C_1(m_1) \times C_2(m_2)$ for transmission. The codebook together with all preselected sequence pairs will then be revealed to all parties. 

\begin{figure}[b]
	\centerline{
		\subfloat[Subcodebook $C_1(1)$ of the Marton coding scheme]{\makebox[7cm][c]{\includegraphics[scale=0.6]{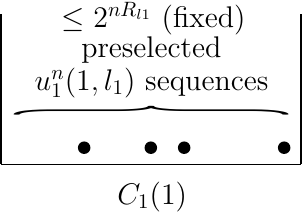}}
		% where an .eps filename suffix will be assumed under latex,
		% and a .pdf suffix will be assumed for pdflatex
		\label{fig:3a}}}
	\centerline{
		%\hfil
		\subfloat[Subcodebook $C_1(1)$ of the Wyner secrecy coding scheme]{\makebox[7cm][c]{\includegraphics[scale=0.6]{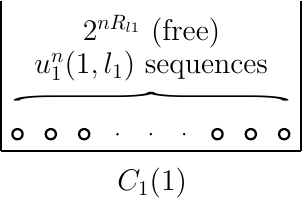}}
		% where an .eps filename suffix will be assumed under latex,
		% and a .pdf suffix will be assumed for pdflatex
		\label{fig:3b}}}
	\caption{Single subcodebook comparison between the Marton coding scheme and the Wyner secrecy coding scheme.}
	\label{fig:3}
\end{figure}

The difference above directly impacts the encoding stage. For this discussion, we assume that $M_1=1$ is sent and consider only the subcodebook $C_1(1)$. We will also assume $2^{nR_{d1}}=2^{nR_{l1}}$. As illustrated in Fig.~\ref{fig:3b}, to transmit the message, Wyner secrecy coding~\cite{cref4} allows the encoder to randomly choose one of the $2^{nR_{l1}}$ $u_1^n(1,l_1)$ sequences in the subcodebook $C_1(1)$ according to the uniform distribution. The large number of remaining $u_1^n(1,l_1)$ sequences can then act as random components to confuse the eavesdropper. This is the essence behind ensuring message protection Wyner secrecy coding.       

In Marton coding~\cite{cref5}, the encoder does not enjoy the same freedom in sequence selection which is observed in Wyner secrecy coding~\cite{cref4}. For instance, assume that we also fix $M_2=1$, from subcodebook $C_1(1)$, the encoder can only choose the $u_1^n(1,l_1)$ sequence that forms a preselected sequence pair with $u_2^n(1,l_2)$. For ease of discussion, we let the sequence be $u_1^n(1,1)$. 

In order for Marton coding~\cite{cref5} to achieve message protection in the same manner as Wyner secrecy coding~\cite{cref4}, we need to ensure that aside from the sequence $u_1^n(1,1)$, we still have additional $u_1^n(1,l_1)$, $l_1\neq 1,$ sequences that can act as random components to confuse the eavesdropper. This is possible if upon considering $C_1(1)$ across all $C_2(m_2)$, we have a large number of or ideally $2^{nR_{l1}}$ distinct $u_1^n(1,l_1)$ sequences that form preselected sequence pairs with $u_2^n(m_2,l_2)$. A sample of this scenario is illustrated in Fig.~\ref{fig:3a}.

\begin{figure}[t]
	\centering
	\includegraphics[scale=0.68]{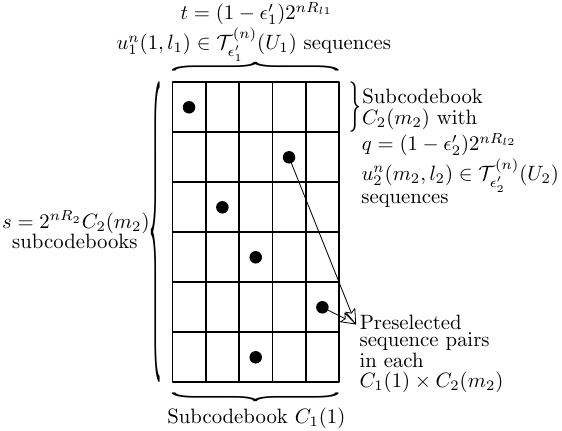}
	\caption{Problem model.}
	\label{fig:4}
\end{figure}

As a continuity of the idea, if we aim to bridge Marton coding~\cite{cref5} and Wyner secrecy coding~\cite{cref4}, the following questions need to be answered: Considering each subcodebook $C_i(m_i)$, $i \in {1,2},$ in $n$ channel uses, are there sufficiently large numbers of distinct $u_i^n(m_i,l_i)$ sequences that form preselected sequence pairs? Is each $u_i^n(m_i,l_i)$ sequence equally likely to be in the preselected sequence pairs? Are there any conditions that we must impose to achieve these requirements?

\subsection{Problem Model}\label{S3C}

Since each subcodebook $C_i(m_i)$ is randomly and independently generated for each message $m_i$, $i=1,2$, it is sufficient for us to consider a single $C_1$ subcodebook. The analysis will apply to all remaining $C_1$ and $C_2$ subcodebooks as well.
%$S(m_i,l_i)$ as $\{u_j^n(m_j,l_j): (u_i^n(m_i,l_i), u_j^n(m_j,l_j)) \in \mathcal{T}^{(n)}_{\epsilon} \text{ for some } m_j \in [1:2^{nR_{j}}] \text{ and } l_j \in [1:2^{nR_{lj}}], \text{ and } ( u_i^n(m_i,l_i), u_2^n(m_j,l_j)  ) \text{ is the selected sequence pair for } (m_i, m_j) \}$
%where $j\triangleq(i \text{ mod }2)+1$, for all $i\in \{1,2\}$
%or
%$B(m1)$ as ${ l1 \in [1:2nRl1] : ( u_1^n(m1,l1), u_2^n(m2,l2)  ) \in T for some m2 \in [1:2nR2] and l2 \in [1:2nRl2] , and ( u_1^n(m1,l1), u_2^n(m2,l2)  ) is the selected sequence pair for (m1, m2) }$
%Note that the bin definitions above are not the same. One is a collection of sequences, and the other a collection of indices.
Our problem model is illustrated in Fig.~\ref{fig:4}. This is part of the Marton coding codebook in Fig.~\ref{fig:6a} after preselecting the sequence pairs. In this case, we consider the $C_1(1)$ subcodebook across all $2^{nR_{2}}$ $C_2(m_2)$ subcodebooks. Note that the product subcodebook $C_1(1)\times C_2(m_2)$ for each $m_2$ contains one preselected sequence pair as required by Marton coding. Our goal here is to identify the number of distinct $u_1^n(1,l_1)$ sequences that form preselected sequence pairs and to determine if each $u_1^n(1,l_1)$ sequence is equally likely to be in the preselected sequence pairs across all $2^{nR_{2}}$ $C_2(m_2)$ subcodebooks. 

In order to simplify our analysis, it is important to take note that our problem model only considers codebook generations which have approximately all $u_1^n$ and $u_2^n$ sequences in their respective typical sets. More precisely, 
the codebook generations of interest satisfy the following properties: 
\begin{itemize}[leftmargin=*]
	\item In every $v=2^{nR_{1}}$ $C_1(m_1)$ subcodebook that contains $t'=2^{nR_{l1}}$ $u_1^n$ sequences, with probability $\geq 1-\epsilon'_1$, we have $t=(1 - \epsilon'_1)2^{nR_{l1}}$ typical $u_1^n$ sequences, i.e., $u_1^n \in \mathcal{T}^{(n)}_{\epsilon'_1}(U_1)$,
	\item In every $s=2^{nR_{2}}$ $C_2(m_2)$ subcodebook that contains $q'=2^{nR_{l2}}$ $u_2^n$ sequences, with probability $\geq 1-\epsilon'_2$, we have $q=(1 - \epsilon'_2)2^{nR_{l2}}$ typical $u_2^n$ sequences, i.e., $u_2^n \in \mathcal{T}^{(n)}_{\epsilon'_2}(U_2)$,
\end{itemize}
where we set $\epsilon'_1 \to 0$ and $\epsilon'_2 \to 0$ as $n \to \infty$. By the union of events bound~\cite{cref6}, with probability of at least $1-\left(v\epsilon'_1 + s\epsilon'_2\right)$, all $C_1(m_1)$ and $C_2(m_2)$ subcodebooks satisfy the two properties above, for all $m_{1}\in [1:v]$ and $m_{2}\in [1:s]$. 

Our definition of the problem model originates from a simple observation. Note that the main criteria for sequence pair preselection in Marton coding requires some $u_1^n$ and $u_2^n$ sequences to be jointly typical, i.e., $(u^n_1,u^n_2)\in \mathcal{T}^{(n)}_{\epsilon'}$. By the properties of jointly typical sequences~\cite{cref6}, this also implies that the preselected $u_1^n$ and $u_2^n$ sequences should be in their respective typical sets, i.e., $u_1^n \in \mathcal{T}^{(n)}_{\epsilon'_1}(U_1)$ and $u_2^n \in \mathcal{T}^{(n)}_{\epsilon'_2}(U_2)$ with $\epsilon'=\min\{\epsilon'_1, \epsilon'_2\}$. By limiting our problem model to such specific codebook generations, we gain the advantage of using existing numerical bounds to aid in our calculations. Although this also indicates that we are only working with a subset of all codebook generations, we will show that it is sufficient to prove the existence of at least one codebook that is both decoding-good and secrecy-good. 

\begin{figure}[t]
	\centering
	\includegraphics[scale=0.68]{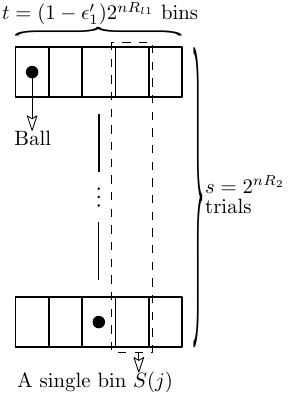}
	\caption{Equivalent ball placement experiment.}
	\label{fig:5}
\end{figure}

\subsection{Equivalent Ball Placement Experiment}\label{S3D}

In order to simplify the notations in our proofs, we temporarily represent our problem model with the ball placement experiment illustrated in Fig.~\ref{fig:5}.  Defining each $u_1^n(1,j)$ position across all $C_2(m_2)$ as a single bin $S(j)$, each product subcodebook $C_1(1)\times C_2(m_2)$ is now equivalent to a collection of $t$ bins, i.e., $\{S(j): j \in [1:t]\}$. Mimicking the outcome of sequence pair preselection in Marton coding, the experiment then involves placing a ball randomly and uniformly into one of the $t$ bins. This process will be repeated independently for $s$ times, corresponding to the independence of sequence pair preselection between each product subcodebook. 

Ideally, in order to achieve a similar degree of randomness provided by Wyner secrecy coding, we would like the $s$ number of balls to be uniformly distributed across the $t$ bins after $s$ trials. Let $p_{Sj}^{(s)}$ be defined as
\begin{flalign}\label{l3} 
p_{Sj}^{(s)} \coloneqq \frac{\text{number of balls in the $j$th bin after $s$ trials}}{s\text{ trials}}\text{.}
\end{flalign}
For all $j\in[1:t]$, we want $p_{Sj}^{(s)} \rightarrow 1/t$, with probability tending to one, as $n\to \infty$.  

Using the ball placement experiment as a reference, we will derive Lemma~\ref{lemma1} and Lemma~\ref{lemma2}. %Besides, since the ball placement experiment is a direct representation of the problem model, Lemma~\ref{lemma1} and Lemma~\ref{lemma2} hold true with probability $\geq 1-\left(v\epsilon'_1 + s\epsilon'_2\right)$.% 
First, we provide a bound on the probability for a ball to be placed in a bin as shown in Lemma~\ref{lemma1}.

%We start our proof by first calculating the probability of each $u_1^n(1,l_1)$ sequence being preselected to form a sequence pair in each $C_1(1) \times C_2(m_2)$. This is shown in Lemma~\ref{lemma1}. 

%Define the subcodebooks $C_i(m_i)$ to consist of $(1 - \epsilon')2^{nR_{li}}$ randomly and independently generated sequences $u_i^n(m_i,l_i) \in \mathcal{T}^{(n)}_{\epsilon'}(U_i)$, for all $i\in\{1,2\}$. Considering the $C_1(1)$ subcodebook across all $C_2(m_2)$ subcodebooks,

\begin{lemma}\label{lemma1}
	 The probability for a ball to be placed in the $j$th bin in the $i$th trial is lower bounded by $\frac{p_{l}}{t}$ and upper bounded by $\frac{p_{u}}{t}$, where $p_{l} = (1-\epsilon')^{2} 2^{-n2\epsilon'\gamma}$, $p_{u} = (1-\epsilon')^{-2} 2^{n2\epsilon'\gamma}$ and	$\gamma =H(U_2|U_1)+H(U_2)$.  
\end{lemma}

\begin{IEEEproof}
	[Proof of Lemma~\ref{lemma1}] Refer to Appendix~\ref{AppA}.
\end{IEEEproof}

%$p_{l} \to 1$, $p_{l} \to 1$, and $p_{u} \to 1$ as $n\to \infty$, for all $i\in\{1,2\}$

%As a consequence of Lemma~\ref{lemma1}, it is also equiprobable for each $u_1^n(1,l_1)$ sequence to be preselected to form a sequence pair in each $C_1(1) \times C_2(m_2)$. 
Using Lemma~\ref{lemma1}, we now provide a bound on the uniformity of $p_{Sj}^{(s)}$ across the $t$ bins after $s$ trials as follows:
\begin{lemma}\label{lemma2}
	For any real number $k>0$, 
	\begin{flalign}
	\mathrm{P}\left(\bigcap\limits_{j=1}^{t} \left|p_{Sj}^{(s)}-\mathrm{E}\left[p_{Sj}^{(s)}\right]\right| < k\sigma_j\right) \geq 1-\frac{t'}{k^2},\label{l2}  
	\end{flalign}
	where $\sigma_j$ is the standard deviation of $p_{Sj}^{(s)}$.
\end{lemma}

\begin{IEEEproof}
	[Proof of Lemma~\ref{lemma2}] Refer to Appendix~\ref{AppB}.
\end{IEEEproof}

\subsection{Sufficient Conditions to Bridge Marton Coding and Wyner Secrecy Coding}\label{S3E}

Reverting from the ball placement experiment to the problem model, we apply Lemma~\ref{lemma2} to all remaining $C_1$ and $C_2$ subcodebooks. With this, we will bound the entropy term $H(L_i|M_i)$ and obtain the sufficient conditions to bridge Marton coding and Wyner secrecy coding. For some choice of $\epsilon'_1$, $\epsilon'_2$ and $k$, Lemma~\ref{lemma3} holds true.

\begin{lemma}\label{lemma3}
	If 
	\begin{flalign}
	R_1>2R_{l2} \text{\space and\space} R_2>2R_{l1}, \label{l3a}  
	\end{flalign}
	then 
	\begin{flalign}
	\mathrm{P}\left(H(L_i|M_i) \geq np_{l} R_{li} - \epsilon_i\right) \geq 1 - \epsilon_{0},\label{l3b}  
	\end{flalign}
	where $p_{l} \to 1$, $\epsilon_i \to 0$ and $\epsilon_{0} \to 0$ as $n\to \infty$, for all $i\in\{1,2\}$. 
\end{lemma}

\begin{IEEEproof}
	[Proof of Lemma~\ref{lemma3}] Refer to Appendix~\ref{AppC}.
\end{IEEEproof}

Satisfying the constraints in (\ref{l3a}) ensures that for all $i\in\{1,2\}$, the probability of each of the following events tends to one as $n\to \infty$:  
\begin{itemize}[leftmargin=*]
	\item The number of distinct $u_i^n(m_i,l_i)$ sequences that form preselected sequence pairs tends to $2^{nR_{li}}$,
	%\item Over a random selection of $M_j$, $j \neq i$, these $u_i^n(m_i,l_i)$ sequences are asymptotically uniformly distributed.
	\item Over a random selection of $M_j$, $j \neq i$, $L_i$ for which $u_i^n(m_i,L_i)$ form a preselected sequence pair is asymptotically uniformly distributed.
\end{itemize}

This shows that asymptotically, with probability tending to one, in each message subcodebook of the Marton coding scheme, approximately all $u_1^n(m_1,l_1)$ and $u_2^n(m_2,l_2)$ sequences are equally likely to form the preselected sequence pairs. This indicates that the random components in Marton coding can play the role  of Wyner random components to confuse the eavesdropper. This in turn shows that the Marton coding scheme alone is capable of providing some basic message protection. 

\subsection{Achievable Individual Secrecy Rate Region}\label{S3F}

The Marton coding scheme achieves the individual secrecy rate region $\mathcal{R}$ in Theorem~\ref{theorem1}.

\begin{theorem}\label{theorem1}
	Using Marton coding, the following individual secrecy rate region is achievable for the two-receiver discrete memoryless broadcast channel with a passive eavesdropper:
	\begin{equation*}\label{eq2}
		\mathcal{R} \triangleq
		\left\{
		\begin{split}
			(R_1,R_2)\\ \in \mathbb{R}_+^2
		\end{split}
		\left\vert
		\begin{split}
			&R_1< \text{min}\{I(U_1;Y_1)-I(U_1;Z),\\
			&2I(U_1,Y_1)-2I(U_1,U_2)\},\\
			&R_2< \text{min}\{I(U_2;Y_2)-I(U_2;Z),\\
			&2I(U_2,Y_2)-2I(U_1,U_2)\},\\
			&R_1+R_2<I(U_1;Y_1)+I(U_2;Y_2)\\
			&-I(U_1;U_2),\\
			&R_1>2I(U_2;Z),\\
			&R_2>2I(U_1;Z),\\
			&R_1+R_2>2I(U_1;U_2),\\
			&\text{over all }p(u_1,u_2)p(x|u_1,u_2)\text{.}\\ 
		\end{split}
		\right.
		\right\}
	\end{equation*}
\end {theorem}

\begin{IEEEproof}
	[Proof of Theorem~\ref{theorem1}] Refer to Appendix~\ref{AppD} for the complete proof.
\end{IEEEproof}
		
\begin{remark}
	The individual secrecy rate region in Theorem~\ref{theorem1} imposes three lower bounds that seemingly bound the region away from the $R_1$ and $R_2$ axes. However, we note that any rate points that are potentially excluded due to these lower bounds can be recovered by noting that if $(R_1,R_2)$ is achievable, so is $(R_1-\lambda_1, R_2-\lambda_2)$ for any $0 \leq \lambda_1 \leq R_1$ and $0 \leq \lambda_2 \leq R_2$. In order to see this, simply inject additional randomness to the transmitted messages, i.e., by taking $M'_{i}=(M_{i},T_{i})$ where $M_{i}\in [1:2^{nR_{i}}]$, $T_{i}\in [1:2^{nR_{ti}}]$ and $R'_{i}=R_{i}+R_{ti}$, for all $i\in\{1,2\}$.
\end{remark}

\section{Discussion and Conclusion}\label{S4}

Ultimately, the results in this paper shows that Marton coding~\cite{cref5} allows us to achieve message protection through individual secrecy. Here, we once again emphasize that the random components in Marton coding $L_1$ and $L_2$ can play a role in message protection at the cost of two additional constraints in (\ref{l3a}). By satisfying these additional constraints, we can guarantee that in each message subcodebook, approximately all $u_1^n(m_1,l_1)$ and $u_2^n(m_2,l_2)$ sequences form the preselected sequence pairs required by Marton coding. This scenario provides us with sufficient $u_1^n(m_1,l_1)$ and $u_2^n(m_2,l_2)$ sequences which can act as random components to confuse the eavesdropper even when the encoder has a strict encoding rule. Our results provides us with an individual secrecy rate region that is potentially larger than those by Chen et al.~\cite{cref1} and Mansour et al.~\cite{Mansour_Schaefer_Boche16}. This can be observed since the secrecy coding schemes by Chen et al.~\cite{cref1} and Mansour et al.~\cite{Mansour_Schaefer_Boche16} do not achieve any positive rate when reduced to Marton coding only. 

In addition, we provide an intuition that with Marton coding alone, joint secrecy might not be achieved. The joint secrecy notion requires the joint information leakage from both message to the eavesdropper to be vanishing~\cite{iref14,iref15}. This requirement is difficult to be satisfied by Marton coding due to the dependency between the $U_1^n$ and $U_2^n$ codewords.

In conclusion, the utilization of the Marton coding scheme~\cite{cref5} to ensure individual secrecy is desirable since it may provide additional message protection, complementing other secrecy techniques. This reduces the complexity of coding schemes, especially when we are dealing with channels with a large number of receivers. It thus opens the opportunity to derive new secrecy coding schemes that performs better than existing ones. Nonetheless, we will further investigate the ease of implementation and compare the performance of this strategy in future works.

\appendices
\section{}\label{AppA}

%\begin{figure}[t]
%	\centering
%	\includegraphics[scale=0.85]{Figures/Sing-CBook-3}
%	\caption{The product subcodebook $C_1(1)\times C_2(1)$.}
%	\label{fig:5}
%\end{figure}

In this section, we will present the proof of Lemma~\ref{lemma1}.

\begin{IEEEproof}
	[Proof of Lemma~\ref{lemma1}] %We start our proof with a quick overview on codebook generation. Fix a pmf $p(u_1,u_2)$. The subcodebook $C_1(1)$ consists of $t'$ randomly and independently generated sequences $u^n_1(1,l_1)$, $l_1\in [1:t']$, each according to $\prod\limits_{i=1}^{n}p_{U_1}(u_{1i})$. Meanwhile, for all $m_2\in [1:s]$, the subcodebook $C_2(m_2)$ consists of $q'$ randomly and independently generated sequences $u^n_2(m_2,l_2)$, $l_2\in [1:q']$, each according to $\prod\limits_{i=1}^{n}p_{U_2}(u_{2i})$. 
	%Satisfying the Marton coding principle, for each product subcodebook $C_1(1)\times C_2(m_2)$, find an index pair $(l_1,l_2)$ such that $(u^n_1(1,l_1),u^n_2(m_2,l_2))\in \mathcal{T}^{(n)}_{\epsilon'}$. If there is more than one such jointly typical pair, the encoder chooses an arbitrary one among those. If no such pair exists, choose $(l_1,l_2)$ uniformly at random from all possible pairs.
	%Recalling our problem model, we have stated that we will consider the message subcodebook $C_1(1)$ which only contains the sequences $u_1^n \in \mathcal{T}^{(n)}_{\epsilon'}(U_1)$ and $u_2^n \in \mathcal{T}^{(n)}_{\epsilon'}(U_2)$.  Note that the main criteria for sequence pair preselection in Marton coding is based upon the fact that some $u_1^n$ and $u_2^n$ sequences are jointly typical. By the properties of jointly typical sequences~\cite{cref6}, this also implies that the preselected $u_1^n$ and $u_2^n$ sequences should be in their respective typical sets. By limiting our problem model to only the sequences $u_1^n \in \mathcal{T}^{(n)}_{\epsilon'}(U_1)$ and $u_2^n \in \mathcal{T}^{(n)}_{\epsilon'}(U_2)$, we can use existing bounds to bound the probability for a ball to be placed in a $S(l_1)$ bin in the ball placement experiment.
	By the properties of jointly typical sequences~\cite{cref6}, given any sequence $u_1^n\in \mathcal{T}^{(n)}_{\epsilon'_1}(U_1)$, we define $p_b$ as the probability that a $u_2^n$ sequence is jointly typical with $u_1^n$ such that 
	\begin{flalign}
		p_b \coloneqq \frac{\left| \mathcal{T}^{(n)}_{\epsilon'}(U_2|u_1^n)\right|}{\left| \mathcal{T}^{(n)}_{\epsilon'}(U_2)\right|}\label{aa1}.
	\end{flalign} 
	Using existing bounds on the terms in the numerator and denominator~\cite{cref6}, we have the lower bound on $p_b$ as
	\begin{flalign}
		p_b &\geq \frac{(1-\epsilon') 2^{n(1-\epsilon')H(U_2|U_1)}}{2^{n(1+\epsilon')H(U_2)}}&\nonumber\\
		&=(1-\epsilon') 2^{-n\left[I(U_1;U_2)+\epsilon'\left( H(U_2|U_1)+H(U_2) \right)\right]}&\nonumber\\
		&\overset{\text{(a)}}{=}(1-\epsilon') 2^{-n\left[I(U_1;U_2)+\epsilon'\gamma\right]}\label{aa2}&
	\end{flalign}
	and the upper bound on $p_b$ as
	\begin{flalign}
		p_b &\leq \frac{ 2^{n(1+\epsilon')H(U_2|U_1)}}{(1-\epsilon')2^{n(1-\epsilon')H(U_2)}}&\nonumber\\
		&=(1-\epsilon')^{-1} 2^{-n\left[I(U_1;U_2)-\epsilon'\left( H(U_2|U_1)+H(U_2) \right)\right]}&\nonumber\\
		&\overset{\text{(b)}}{=}(1-\epsilon')^{-1} 2^{-n\left[I(U_1;U_2)-\epsilon'\gamma\right]}\label{aa3},&
	\end{flalign}
	where (a) and (b) follows by defining $\gamma \coloneqq H(U_2|U_1)+H(U_2)$.
	
	Now, referring to the ball placement experiment, for all $i \in [1:s]$ and $j \in [1:t]$, we define $A_{i,j}$ as the random variable representing the number of jointly typical sequence pairs (before preselection) in the $j$th bin in the $i$th trial. The expected value of $A_{i,j}$, $\mathrm{E}[A_{i,j}]$ is calculated as
	\begin{flalign}
		\mathrm{E}[A_{i,j}] &= \sum\limits_{u_2^n \in S(j)} p_b&\nonumber\\
		&=p_b q\label{aa4}.&
	\end{flalign}
	Upon substituting (\ref{aa2}) into (\ref{aa4}), we have the lower bound on $\mathrm{E}[A_{i,j}]$ as
	\begin{flalign}
		\mathrm{E}[A_{i,j}] \geq (1-\epsilon') 2^{-n\left[I(U_1;U_2)+\epsilon'\gamma\right]}q \label{aa5}.
	\end{flalign}
	Upon substituting (\ref{aa3}) into (\ref{aa4}), we have the upper bound on $\mathrm{E}[A_{i,j}]$ as
	\begin{flalign}
		\mathrm{E}[A_{i,j}] \leq (1-\epsilon')^{-1} 2^{-n\left[I(U_1;U_2)-\epsilon'\gamma\right]}q \label{aa6}.
	\end{flalign}
	
	We also define $B_{i}$ as the random variable representing the number of jointly typical sequence pairs (before preselection) in all $t$ bins in the $i$th trial. The expected value of $B_{i}$, $\mathrm{E}[B_{i}]$ is
	\begin{flalign}
		\mathrm{E}[B_{i}] &= t \left(\sum\limits_{u_2^n \in S(j)} p_b\right)&\nonumber\\
		&=p_b qt\label{aa7}.&
	\end{flalign}
	Upon substituting (\ref{aa2}) into (\ref{aa7}), we have the lower bound on $\mathrm{E}[B_{i}]$ as
	\begin{flalign}
		\mathrm{E}[B_{i}] \geq (1-\epsilon') 2^{-n\left[I(U_1;U_2)+\epsilon'\gamma\right]}qt \label{aa8}.
	\end{flalign}
	Upon substituting (\ref{aa3}) into (\ref{aa7}), we have the upper bound on $\mathrm{E}[B_{i}]$ as
	\begin{flalign}
		\mathrm{E}[B_{i}] \leq (1-\epsilon')^{-1} 2^{-n\left[I(U_1;U_2)-\epsilon'\gamma\right]}qt \label{aa9}.
	\end{flalign} 
	
	Next, we define the indicator random variable $G_{i,j}$ such that 
	\begin{equation*}
		G_{i,j} \coloneqq
		\left\{
		\begin{split}
		&1 \text{\quad if a ball is placed in the $j$th bin in the $i$th trial,}\\ 
		&0 \text{\quad otherwise.}
		\end{split}
		\right.
	\end{equation*}
	We are interested in calculating the upper and lower bound on $\mathrm{P}(G_{i,j}=1)$ which can be defined as 
	\begin{flalign}
		\mathrm{P}(G_{i,j}=1)\coloneqq \frac{\mathrm{E}[A_{i,j}]}{\mathrm{E}[B_{i}]}\label{aa10}.
	\end{flalign}
	Upon substituting (\ref{aa5}) and (\ref{aa9}) and into (\ref{aa10}), we have the lower bound on $\mathrm{P}(G_{i,j}=1)$ as
	\begin{flalign}
	\mathrm{P}(G_{i,j}=1) &\geq \frac{(1-\epsilon') 2^{-n\left[I(U_1;U_2)+\epsilon'\gamma\right]}q}{(1-\epsilon')^{-1} 2^{-n\left[I(U_1;U_2)-\epsilon'\gamma\right]}qt}&\nonumber\\
	&=\frac{(1-\epsilon')^{2} 2^{-n2\epsilon'\gamma}}{t}&\nonumber\\
	&\overset{\text{(c)}}{=} \frac{p_l}{t} \label{aa11},&
	\end{flalign}
	where (c) follows by defining $p_{l} \coloneqq (1-\epsilon')^{2} 2^{-n2\epsilon'\gamma}$. Upon substituting (\ref{aa6}) and (\ref{aa8}) into (\ref{aa10}), we have the upper bound on $\mathrm{P}(G_{i,j}=1)$ as
	\begin{flalign}
	\mathrm{P}(G_{i,j}=1) &\leq \frac{(1-\epsilon')^{-1} 2^{-n\left[I(U_1;U_2)-\epsilon'\gamma\right]}q}{(1-\epsilon') 2^{-n\left[I(U_1;U_2)+\epsilon'\gamma\right]}qt}&\nonumber\\
	&=\frac{(1-\epsilon')^{-2} 2^{n2\epsilon'\gamma}}{t}&\nonumber\\
	&\overset{\text{(d)}}{=} \frac{p_u}{t} \label{aa12},&
	\end{flalign} 
	where (d) follows by defining $p_{u} \coloneqq (1-\epsilon')^{-2} 2^{n2\epsilon'\gamma}$. This completes the proof of Lemma~\ref{lemma1}.
\end{IEEEproof}

\section{}\label{AppB}

In this section, we will present the proof of Lemma~\ref{lemma2}.

\begin{IEEEproof}
	[Proof of Lemma~\ref{lemma2}] %Continuing from Appendix~\ref{AppA}, with probability $\geq 1-(s + 1)\epsilon'$, we will consider the message subcodebook $C_1(1)$ which only contains the sequences $u_1^n \in \mathcal{T}^{(n)}_{\epsilon'}(U_1)$ and $u_2^n \in \mathcal{T}^{(n)}_{\epsilon'}(U_2)$. 
	Corresponding to the numerator in definition (\ref{l3}), we define $J_j$ as the random variable representing the number of balls in the $j$th bin after $s$ trials such that 
	\begin{flalign}
		J_j \coloneqq \sum_{i=1}^{s} G_{i,j}\label{cc1},
	\end{flalign}
	where $G_{i,j}$ is the indicator random variable defined in Appendix~\ref{AppA}.

	Now, we calculate the expected value of $J_j$.
	\begin{flalign}
		\mathrm{E}[J_j]&\overset{\text{(a)}}{=} \mathrm{E}\left[\sum_{i=1}^{s} G_{i,j}\right]&\nonumber\\
		&\overset{\text{(b)}}{=} \quad \sum_{i=1}^{s}\mathrm{E}[G_{i,j}]&\nonumber\\
		&= \quad \sum_{i=1}^{s} \mathrm{P}(G_{i,j}=1)&\nonumber\\
		&\overset{\text{(c)}}{\geq} \quad \sum_{i=1}^{s} \frac{p_{l}}{t}&\nonumber\\
		&=\frac{p_{l} s}{t}\label{cc2},&
	\end{flalign}
	where (a) follows from substituting (\ref{cc1}), (b) follows from the linearity of expected value and (c) follows from Lemma~\ref{lemma1}. This gives us the lower bound on the expected value of $J_j$.

	Next, we calculate the variance of $J_j$ which is defined as 
	\begin{flalign}
		\mathrm{Var}[J_j]\coloneqq\mathrm{E}\left[J_j^2\right]-\left(\mathrm{E}[J_j]\right)^2\label{cc5}.
	\end{flalign}
	Calculating the first term $\mathrm{E}\left[J_j^2\right]$, we have
	\begin{flalign}
		\mathrm{E}\left[J_j^2\right]&\overset{\text{(d)}}{=}\mathrm{E}\left[\left(\sum_{i=1}^{s}G_{i,j}\right)^2\right]&\nonumber\\
		&=\mathrm{E}\left[\left(\sum_{i=1}^{s}G_{i,j}\right)\left(\sum_{k=1}^{s}G_{k,j}\right)\right]&\nonumber\\
		&=\mathrm{E}\left[\left(\sum_{i=1}^{s}G_{i,j}\right)\left(G_{i,j}+\sum_{k\neq i}^{s}G_{k,j}\right)\right]&\nonumber\\
		&=\mathrm{E}\left[\sum_{i=1}^{s}G_{i,j}^2+\sum_{i=1}^{s}\sum_{k\neq i}^{s}G_{i,j}G_{k,j}\right]&\nonumber\\
		&\overset{\text{(e)}}{=}\mathrm{E}\left[\sum_{i=1}^{s}G_{i,j}^2\right]+\mathrm{E}\left[\sum_{i=1}^{s}\sum_{k\neq i}^{s}G_{i,j}G_{k,j}\right]&\nonumber\\
		&\overset{\text{(f)}}{=}\sum_{i=1}^{s}\mathrm{E}\left[G_{i,j}^2\right]+\sum_{i=1}^{s}\sum_{k\neq i}^{s}\mathrm{E}\left[G_{i,j}G_{k,j}\right]\label{cc6},&
	\end{flalign}
	where (d) follows from substituting (\ref{cc1}) while both (e) and (f) follows from the linearity of expected value. Moving on, we need to calculate $\mathrm{E}\left[G_{i,j}^2\right]$ and $\mathrm{E}\left[G_{i,j}G_{k,j}\right]$. Since $G_{i,j} \in \{0,1\}$,
	\begin{flalign}
		\mathrm{E}\left[G_{i,j}^2\right]&=\mathrm{E}\left[G_{i,j}\right]&\nonumber\\
		&=\mathrm{P}(G_{i,j}=1)&\nonumber\\
		&\overset{\text{(g)}}{\leq}\frac{p_{u}}{t}\label{cc7},&
	\end{flalign}
	where (g) follows from Lemma~\ref{lemma1}. Since each trial is independent,  
	\begin{flalign}
		\mathrm{E}\left[G_{i,j}G_{k,j}\right]
		&=\mathrm{E}\left[G_{i,j}\right] \mathrm{E}\left[G_{k,j}\right]&\nonumber\\
		&\leq\frac{p_{u}^2}{t^2}\label{cc8}.&
	\end{flalign}
	Upon substituting (\ref{cc7}) and (\ref{cc8}) into (\ref{cc6}), we have 
	\begin{flalign}
		\mathrm{E}\left[J_j^2\right]
		&\leq \sum_{i=1}^{s}\frac{p_{u}}{t}+\sum_{i=1}^{s}\sum_{k\neq i}^{s}\frac{p_{u}^2}{t^2}&\nonumber\\
		&=\frac{p_{u}s}{t}+\frac{s(s-1)p_{u}^2}{t^2}\label{cc9}.&
	\end{flalign}
	Upon substituting (\ref{cc2}) and (\ref{cc9}) into (\ref{cc5}), we have  
	\begin{flalign}
		\mathrm{Var}[J_j]
		&\leq \frac{p_{u}s}{t}+\frac{p_{u}^2s(s-1)}{t^2}-\frac{p_{l}^2 s^2}{t^2}&\nonumber\\
		&=\frac{p_{u}s}{t}-\frac{p_{u}^2s}{t^2}+\frac{p_{u}^2 s^2}{t^2}-\frac{p_{l}^2 s^2}{t^2}\label{cc10},&
	\end{flalign}
	giving us the upper bound on the variance of $J_j$.
	
	With the bounds on $\mathrm{E}[J_j]$ and $\mathrm{Var}[J_j]$, we proceed to calculate the bounds on the expected value and the variance of $p_{Sj}^{(s)}$, i.e., $\mathrm{E}\left[p_{Sj}^{(s)}\right]$ and $\mathrm{Var}\left[p_{Sj}^{(s)}\right]$. As a consequence of the definitions (\ref{l3}) and (\ref{cc1}), we have
	\begin{flalign}
		\mathrm{E}\left[p_{Sj}^{(s)}\right]
		&=\frac{\mathrm{E}[J_j]}{s}&\nonumber\\
		&\overset{\text{(h)}}{\geq}\frac{p_{l}}{t}\label{cc11},&
	\end{flalign}
	where (h) follows from substituting (\ref{cc2}). Similarly, 
	\begin{flalign}
		\mathrm{Var}\left[p_{Sj}^{(s)}\right]
		&=\frac{\mathrm{Var}[J_j]}{s^2}&\nonumber\\
		&\overset{\text{(i)}}{\leq}\frac{p_{u}}{st}-\frac{p_{u}^2}{st^2}+\frac{p_{u}^2}{t^2}-\frac{p_{l}^2}{t^2}\label{cc12},&
	\end{flalign}
	where (i) follows from substituting (\ref{cc10}).
	
	Now, we bound $\mathrm{P}\left(\bigcap\limits_{j=1}^{t} \left|p_{Sj}^{(s)}-\mathrm{E}\left[p_{Sj}^{(s)}\right]\right| < k_j\sigma_j\right)$, where $\sigma_j= \sqrt{\mathrm{Var}\left[p_{Sj}^{(s)}\right]}$ is the standard deviation of $p_{Sj}^{(s)}$ for any real number $k_j>0$. 
	\begin{flalign}
		&\mathrm{P}\left(\bigcap\limits_{j=1}^{t} \left|p_{Sj}^{(s)}-\mathrm{E}\left[p_{Sj}^{(s)}\right]\right| < k_j\sigma_j\right)\nonumber&\\  
		&=1-\mathrm{P}\left(\bigcup\limits_{j=1}^{t} \left|p_{Sj}^{(s)}-\mathrm{E}\left[p_{Sj}^{(s)}\right]\right| \geq k_j\sigma_j\right)\nonumber&\\ 
		&\overset{\text{(j)}}{\geq}1-\mathrm{P}\left(\bigcup\limits_{j=1}^{t} \left|p_{Sj}^{(s)}-\mathrm{E}\left[p_{Sj}^{(s)}\right]\right| \geq k\sigma_j\right)\nonumber&\\ 
		&\overset{\text{(k)}}{\geq}1-\frac{t}{k^2}\nonumber&\\
		&\geq 1-\frac{t'}{k^2},&\label{cc13}  
	\end{flalign}
	where (j) follows by defining $k \coloneqq \min\limits_{j} k_j$ and (k) follows from the union of events bound and Chebyshev inequality~\cite{cref6}. This completes the proof of Lemma~\ref{lemma2}.
\end{IEEEproof}

\section{}\label{AppC}

In this section, we will present the proof of Lemma~\ref{lemma3}.

\begin{IEEEproof}
	[Proof of Lemma~\ref{lemma3}] Before proceeding, notice that several error events can be identified from our discussions so far:
	\begin{itemize}[leftmargin=*]
		\item From Section~\ref{S3C}, $u_1^n \notin \mathcal{T}^{(n)}_{\epsilon'_1}(U_1)$ and $u_2^n \notin \mathcal{T}^{(n)}_{\epsilon'_2}(U_2)$ with probability $v\epsilon'_1 + s\epsilon'_2$, 
		\item From Lemma~\ref{lemma2}, $L_1$ is not uniformly distributed after sequence pair preselection with probability $\frac{t'}{k^2}$. 
	\end{itemize}
	By the union of events bound~\cite{cref6}, we define $\epsilon_{0}$ as the upper bound on the mentioned probabilities of error such that $\epsilon_{0} \coloneqq v\epsilon'_1 + s\epsilon'_2 + \frac{t'}{k^2}$. 
	
	Now, we bound the entropy terms $H(L_{1}|M_{1})$ and $H(L_{2}|M_{2})$. By the definition of entropy, 
	\begin{flalign}
		&H(L_1|M_1)&\nonumber\\
		&=\sum_{m_1\in\mathcal{M}_1}p_{M_1}(m_1)H(L_1|M_1=m_1)&\nonumber\\
		&=-\sum_{m_1\in\mathcal{M}_1}p_{M_1}(m_1)\sum_{l_1\in\mathcal{L}_1}p_{L_1|M_1}(l_1|m_1)\log p_{L_1|M_1}(l_1|m_1)&\nonumber\\
		&\overset{\text{(a)}}{=}-\sum_{m_1=1}^{2^{nR_{1}}}\frac{1}{2^{nR_{1}}}\sum_{l_1\in\mathcal{L}_1}p_{L_1|M_1}(l_1|m_1)\log p_{L_1|M_1}(l_1|m_1)&\nonumber\\
		&=-\sum_{l_1\in\mathcal{L}_1}p_{L_1|M_1}(l_1|m_1)\log p_{L_1|M_1}(l_1|m_1)&\nonumber\\
		&\overset{\text{(b)}}{=}-\sum_{j=1}^{t'} p_{Sj}^{(s)}\log p_{Sj}^{(s)}&\nonumber\\
		&\geq -\sum_{j=1}^{t} p_{Sj}^{(s)}\log p_{Sj}^{(s)},&\nonumber
	\end{flalign}
	where (a) follows from the fact that $M_1$ is generated over a uniform distribution and (b) follows from the fact that the definition of $p_{Sj}^{(s)}$ in (\ref{l3}) maps directly to $p_{L_1|M_1}(l_1|m_1)$ when we consider the message subcodebook $C_1(m_1)$.
	
	As a consequence of Lemma~\ref{lemma2}, with probability $\geq 1-\epsilon_{0}$,
	\begin{flalign}	
		&-\sum_{j=1}^{t} p_{Sj}^{(s)}\log p_{Sj}^{(s)}&\nonumber\\
		&\geq -\sum_{j=1}^{t} \left(\mathrm{E}\left[p_{Sj}^{(s)}\right]- k\sigma_j\right)\log \left(\mathrm{E}\left[p_{Sj}^{(s)}\right]- k\sigma_j\right)&\nonumber\\
		&\overset{\text{(c)}}{\geq} -t \left(\frac{p_{l}}{t} - k\sigma\right)\log \left(\frac{p_{l}}{t} - k\sigma\right)&\nonumber\\
		&= p_{l}\log t - (p_{l}-kt\sigma)\log (p_{l}-kt\sigma) - kt\sigma \log t&\nonumber\\
		&= p_{l}\log t' + p_{l}\log (1-\epsilon') - (p_{l}-kt\sigma)\log (p_{l}-kt\sigma)&\nonumber\\
		&\quad - kt\sigma \log t&\nonumber\\
		&\overset{\text{(d)}}{=}np_{l}R_{l1} - \epsilon_1,&\label{bb1}
	\end{flalign}
	where (c) follows from substituting (\ref{cc11}) and from (\ref{cc12}) in which we define $\sigma \coloneqq \left(\frac{p_{u}}{st}-\frac{p_{u}^2}{st^2}+\frac{p_{u}^2}{t^2}-\frac{p_{l}^2}{t^2}\right)^{\frac{1}{2}}$ as the upper bound on $\sigma_j$ and (d) follows by defining $\epsilon_1 \coloneqq -p_{l}\log (1-\epsilon') + (p_{l}-kt\sigma)\log (p_{l}-kt\sigma) + kt\sigma \log t$. 
	
	Note that we want $p_{l} \to 1$, $\epsilon_1 \to 0$ and $\epsilon_{0}\to 0$ as $n\to \infty$. Choose $\epsilon'_1=2^{-n\left(R_{1} + \lambda\right)}$, $\epsilon'_2=2^{-n\left(R_{2} + \lambda\right)}$ and $k=2^{n\left(\frac{R_{l1}}{2}+\frac{\lambda}{2}\right)}$ for an arbitrarily small $\lambda>0$. Suppose $\epsilon'=\min\{\epsilon'_1, \epsilon'_2\}=\epsilon'_2$, we have 
	\begin{flalign}
		p_{l}
		&=(1-\epsilon')^{2} 2^{-n2\epsilon'\gamma}&\nonumber\\
		&=(1-\epsilon')^{2} 2^{-n2\gamma 2^{-n\left(R_{2} + \lambda\right)}}.&\label{bb2}
	\end{flalign}
	From (\ref{bb2}), we calculate the limit of $p_{l}$ as $n\to \infty$ as follows:
	\begin{flalign}
		&\lim\limits_{n\to \infty}(1-\epsilon')^{2} 2^{-n2\gamma 2^{-n\left(R_{2} + \lambda\right)}}&\nonumber\\
		&=\lim\limits_{n\to \infty}(1-\epsilon')^{2}\cdot \lim\limits_{n\to \infty}2^{-n2\gamma 2^{-n\left(R_{2} + \lambda\right)}}&\nonumber\\
		&\overset{\text{(e)}}{=} \lim\limits_{n\to \infty}2^{-n2\gamma 2^{-n\left(R_{2} + \lambda\right)}},\label{bb3}&
	\end{flalign}
	where (e) follows since $\lim\limits_{n\to \infty}\left(1-\epsilon'\right)^{2}=1$. From (\ref{bb3}), letting
	\begin{flalign}
		\lim\limits_{n\to \infty} 2^{-n2\gamma 2^{-n\left(R_{2} + \lambda\right)}}=e^{D}\label{bb4},
	\end{flalign}
	gives us 
	\begin{flalign}
		D&=\lim\limits_{n\to \infty} \ln\left[2^{-n2\gamma 2^{-n\left(R_{2} + \lambda\right)}} \right]&\nonumber\\
		&=\lim\limits_{n\to \infty} -n2\gamma 2^{-n\left(R_{2} + \lambda\right)}\ln(2)&\nonumber\\
		&=-2\gamma\ln(2) \lim\limits_{n\to \infty} n 2^{-n\left(R_{2} + \lambda\right)}&\nonumber\\
		&=-2\gamma\ln(2) \lim\limits_{n\to \infty}\frac{n}{2^{n\left(R_{2} + \lambda\right)}}&\nonumber\\
		&\overset{\text{(f)}}{=} -2\gamma\ln(2) \lim\limits_{n\to \infty} \frac{\frac{d}{dn}n}{\frac{d}{dn}2^{n\left(R_{2} + \lambda\right)}}&\nonumber\\	
		&= -2\gamma\ln(2) \lim\limits_{n\to \infty} \frac{1}{\left(R_{2} + \lambda\right)2^{n\left(R_{2} + \lambda\right)}\ln(2)}&\nonumber\\
		&=0\label{bb5},
	\end{flalign}
	where (f) follows from applying L'H$\hat{\text{o}}$pital's Rule. Upon substituting (\ref{bb5}) into (\ref{bb4}),
	\begin{flalign}
		\lim\limits_{n\to \infty} p_l = 1\label{bb6}.
	\end{flalign}   
	Using a similar approach, we can also show that 
	\begin{flalign}
		\lim\limits_{n\to \infty}p_{u}=1\label{bb7}.
	\end{flalign}
	For $\epsilon_{0}$, we have
	\begin{flalign}
		\epsilon_{0}
		&=v\epsilon'_1 + s\epsilon'_2 +\frac{t'}{k^2}&\nonumber\\
		&=2^{nR_{1}} 2^{-n\left(R_{1} + \lambda\right)} + 2^{nR_{2}} 2^{-n\left(R_{2} + \lambda\right)} + 2^{nR_{l1}} 2^{-n2\left(\frac{R_{l1}}{2}+\frac{\lambda}{2}\right)}&\nonumber\\
		&=3\cdot 2^{-n\lambda}.&\label{bb8}
	\end{flalign}
	From (\ref{bb8}), it is straightforward that $\epsilon_{0}\to 0$ as $n\to \infty$. For $\epsilon_1$, we have
	\begin{flalign}
		\epsilon_1
		&=-p_{l}\log (1-\epsilon') + (p_{l}-kt\sigma)\log (p_{l}-kt\sigma) + kt\sigma \log t&\nonumber\\
		&\overset{\text{(g)}}{\leq} -p_{l}\log (1-\epsilon') + kt\sigma \log t&\nonumber\\
		&= -p_{l}\log (1-\epsilon') + kt\sigma \log t' + kt\sigma \log (1 - \epsilon')&\nonumber\\
		&\overset{\text{(h)}}{\leq} -p_{l}\log (1-\epsilon') + kt\sigma \log t'&\nonumber\\
		&= -p_{l}\log (1-\epsilon')&\nonumber\\
		&\quad + \left(nR_{l1} \right)\left(\frac{k^2 p_{u} t}{s}-\frac{k^2 p_{u}^2}{s}+k^2 p_{u}^2-k^2 p_{l}^2\right)^{\frac{1}{2}},&\label{bb9}
	\end{flalign}
	where (g) follows since $p_{l}\log (1-\epsilon') \leq 0$, $0 \leq p_{l}-kt\sigma \leq 1$ and $(p_{l}-kt\sigma)\log (p_{l}-kt\sigma) \leq 0$ and (h) follows since $kt\sigma \log (1 - \epsilon') \leq 0$. From (\ref{bb9}), we calculate the limit of the two terms as $n\to \infty$ as follows:
	\begin{flalign}
		&\lim\limits_{n\to \infty}-p_{l}\log (1-\epsilon')&\nonumber\\
		&=-\lim\limits_{n\to \infty}p_{l}\cdot \lim\limits_{n\to \infty}\log (1-\epsilon')&\nonumber\\
		&\overset{\text{(i)}}{=} -1(0)&\nonumber\\
		&= 0,\label{bb10}&
	\end{flalign}
	where (i) follows since $\lim\limits_{n\to \infty}\left(1-\epsilon'\right)=1$ and by substituting (\ref{bb6});
	\begin{flalign}
		&\lim\limits_{n\to \infty}\left(nR_{l1} \right)\left(\frac{k^2 p_{u} t}{s}-\frac{k^2 p_{u}^2}{s}+k^2 p_{u}^2-k^2 p_{l}^2\right)^{\frac{1}{2}}&\nonumber\\
		&=\left[\lim\limits_{n\to \infty}\frac{n^2R_{l1}^2}{\left(\frac{k^2 p_{u} t}{s}-\frac{k^2 p_{u}^2}{s}+k^2 p_{u}^2-k^2 p_{l}^2\right)^{-1}}\right]^{\frac{1}{2}}&\nonumber\\
		&\overset{\text{(j)}}{=}\left[\lim\limits_{n\to \infty}\frac{n^2R_{l1}^2}{\left(\frac{k^2 t}{s}-\frac{k^2}{s}\right)^{-1}}\right]^{\frac{1}{2}}&\nonumber\\
		&< \left[\lim\limits_{n\to \infty}\frac{n^2R_{l1}^2}{\left(\frac{k^2 t}{s}\right)^{-1}}\right]^{\frac{1}{2}}&\nonumber\\
		&=\lim\limits_{n\to \infty}\frac{nR_{l1}}{2^{-n\left(R_{l1}-\frac{R_{2}}{2}+\frac{\lambda}{2}\right)}}\cdot \lim\limits_{n\to \infty}\left(1-\epsilon'\right)^{\frac{1}{2}}&\nonumber\\
		&\overset{\text{(k)}}{=} \lim\limits_{n\to \infty}\frac{\frac{d}{dn}nR_{l1}}{\frac{d}{dn}2^{-n\left(R_{l1}-\frac{R_{2}}{2}+\frac{\lambda}{2}\right)}}&\nonumber\\
		&= \lim\limits_{n\to \infty}\frac{R_{l1}}{-\left(R_{l1}-\frac{R_{2}}{2}+\frac{\lambda}{2}\right) 2^{-n\left(R_{l1}-\frac{R_{2}}{2}+\frac{\lambda}{2}\right)} \ln 2},\label{bb11}&
	\end{flalign}
	where (j) follows by substituting (\ref{bb6}) and (\ref{bb7}) and (k) follows since $\lim\limits_{n\to \infty}\left(1-\epsilon'\right)^{\frac{1}{2}}=1$ and from applying L'H$\hat{\text{o}}$pital's Rule. Hence, if $R_2>2R_{l1}$, the limit in (\ref{bb11}) equals to zero. Combining this fact with (\ref{bb10}), we show that $\epsilon_1 \to 0$ as $n\to \infty$. Similarly, we show that if $R_1>2R_{l2}$, then $\mathrm{P}\left(H(L_{2}|M_{2}) \geq np_{l} R_{l2} - \epsilon_2\right)  \geq 1-\epsilon_{0}$ where $p_{l} \to 1$, $\epsilon_2 \to 0$ and $\epsilon_{0}\to 0$ as $n\to \infty$. This completes the proof of Lemma~\ref{lemma3}.
\end{IEEEproof}

\section{}\label{AppD}

In this section, we will present the proof of Theorem~\ref{theorem1}.

\begin{IEEEproof}
	[Proof of Theorem~\ref{theorem1}] \textbf{Sequence generation.} Fix a pmf $p(u_1,u_2)p(x|u_1,u_2)$. For each message $m_{1}\in [1:2^{nR_{1}}]$, generate a subcodebook $C_1(m_1)$ consisting of $2^{nR_{l1}}$ randomly and independently generated sequences $u^n_1(m_{1},l_1)$, $l_1\in [1:2^{nR_{l1}}]$, each according to $\prod\limits_{i=1}^{n}p_{U_1}(u_{1i})$. Once again, for each message $m_{2}\in [1:2^{nR_{2}}]$, generate a subcodebook $C_2(m_2)$ consisting of $2^{nR_{l2}}$ randomly and independently generated sequences $u^n_2(m_{2},l_2)$, $l_2\in [1:2^{nR_{l2}}]$, each according to $\prod\limits_{i=1}^{n}p_{U_2}(u_{2i})$. Note that since we would like to ensure individual secrecy with this codebook using Wyner secrecy coding~\cite{cref4}, the constraints in (\ref{l3a}) need to be satisfied.
		
	\textbf{Sequence pair preselection.} For each message pair $(m_1,m_2)$, find an index pair $(l_1,l_2)$ such that $(u^n_1(m_{1},l_1),u^n_2(m_{2},l_2))\in \mathcal{T}^{(n)}_{\epsilon'}$. If there is more than one such jointly typical pair, the encoder chooses an arbitrary one among those. If no such pair exists, choose $(l_1,l_2)$ uniformly at random from all possible pairs. Lastly, given the preselected sequence pair $(u^n_1,u^n_2)$, randomly generate the codeword $X^n(m_1,m_2)\sim\prod\limits_{i=1}^{n}p_{X|U_1,U_2}(x_i|u_{1i}(m_{1},l_1),u_{2i}(m_{2},l_2))$. This codebook is revealed to all parties (including the eavesdropper).
	
	\textbf{Encoding.} To send the message pair $(m_1,m_2)$, the encoder transmits $x^n(m_1,m_2)$. 
	
	\textbf{Decoding.} Let $\epsilon>\epsilon'$. Receiver 1 declares that $\hat{m}_1$ is sent if it is the unique message such that $(u^n_1(m_1,l_1),y^n_1)\in\mathcal{T}^{(n)}_{\epsilon}$ for some $l_1$; otherwise it declares an error. Similarly, receiver 2 declares that $\hat{m}_2$ is sent if it is the unique message such that $(u^n_2(m_2,l_2),y^n_2)\in\mathcal{T}^{(n)}_{\epsilon}$ for some $l_2$; otherwise it declares an error.
	
	\textbf{Analysis of the probability of error.} Assume without loss of generality that the transmitted messages are equal to one and $(l_1,l_2)=(l'_1,l'_2)$. Receiver 1 makes an error only if one or more of the error events (\ref{dd2})--(\ref{dd4}) occur. On the other hand, receiver 2 makes an error only if one or more of the error events (\ref{dd2}), (\ref{dd5}) and (\ref{dd6}) occur.
	\begin{flalign}
		\mathcal{E}_0&:(U_1^n(1,l_1),U_2^n(1,l_2))\notin \mathcal{T}^{(n)}_{\epsilon'} \text{ for all } l_1 \text{ and } l_2\label{dd2}&\\
		\mathcal{E}_{11}&:(U_1^n(1,l'_1),Y_1^n)\notin \mathcal{T}^{(n)}_{\epsilon}\label{dd3}&\\
		\mathcal{E}_{12}&:(U_1^n(m_{1},l_1),Y_1^n)\in \mathcal{T}^{(n)}_{\epsilon} \text{ for some } m_{1}\neq 1 \text{ and } l_1\label{dd4}&\\
		\mathcal{E}_{21}&:(U_2^n(1,l'_2),Y_2^n)\notin \mathcal{T}^{(n)}_{\epsilon}\label{dd5}&\\
		\mathcal{E}_{22}&:(U_2^n(m_{2},l_2),Y_1^n)\in \mathcal{T}^{(n)}_{\epsilon} \text{ for some } m_{2}\neq 1 \text{ and } l_2\label{dd6}&
	\end{flalign}
	
	Considering the error events, we apply the mutual covering lemma and the packing lemma to show that if
	\begin{flalign}
		R_{l1}+R_{l2}&>I(U_1;U_2),\label{dd7}\\
		R_1+R_{l1}&<I(U_1;Y_1),\label{dd8}\\
		R_2+R_{l2}&<I(U_2;Y_2),\label{dd9}
	\end{flalign}
	then the average probability of decoding error over all codebook generations $\mathrm{P}(\mathcal{E}) \to 0$ as $n\to \infty$.
	
	\textbf{Analysis of individual secrecy.} In order to ensure the individual secrecy of both messages, we need to satisfy (\ref{sec}), i.e., show that $I(M_1;Z^n)\leq n\tau$ and $I(M_2;Z^n)\leq n\tau$.
	
	For the individual secrecy of $M_1$, we have
	\begin{flalign}
		&I(M_1;Z^n)&\nonumber\\
		&=H(M_{1})-H(M_{1}|Z^n)&\nonumber\\
		&=nR_{1}-H(M_{1},L_{1}|Z^n)+H(L_{1}|Z^n,M_{1}).&\label{dd10}
	\end{flalign}
	
	We establish a lower bound on the equivocation term $H(M_{1},L_{1}|Z^n)$ in (\ref{dd10}).
	\begin{flalign}
		&H(M_{1},L_{1}|Z^n)\nonumber&\\
		&=H(M_{1},L_{1}) - I(M_{1},L_{1};Z^n)\nonumber&\\
		&=H(M_{1}) + H(L_{1}|M_{1}) - I(M_{1},L_{1};Z^n).\nonumber&
	\end{flalign}
	
	As a consequence of Lemma~\ref{lemma3}, over all codebook generations, with probability $\geq 1-\epsilon_{0}$, where $\epsilon_{0} \to 0$ as $n\to \infty$,
	\begin{flalign}
		&H(M_{1}) + H(L_{1}|M_{1}) - I(M_{1},L_{1};Z^n)\nonumber&\\
		&\geq n[R_{1}+R_{l1}] - I(M_{1},L_{1};Z^n) - \epsilon_1\nonumber&\\
		&=n[R_{1}+R_{l1}] - I(U^n_1,M_{1},L_{1};Z^n) - \epsilon_1\nonumber&\\
		&\overset{\text{(a)}}{=}n[R_{1}+R_{l1}] - I(U^n_1;Z^n) - \epsilon_1\nonumber&\\
		&\overset{\text{(b)}}{=}n[R_{1}+R_{l1}] - nI(U_1;Z) - \epsilon_1,\label{dd11}&
	\end{flalign}
	where (a) follows since $(M_{1},L_{1})\rightarrow U^n_1\rightarrow Z^n$ forms a Markov chain and (b) follows since by construction $p(u_1^n,z^n) = \prod\limits_{i=1}^{n}p_{U_1,Z}(u_{1i},z_i)$.
	
	We establish an upper bound on the equivocation term $H(L_{1}|Z^n,M_{1})$ in (\ref{dd10}). By~\cite[Lemma 22.1]{cref6}, 
	%\vspace{-0.75pt}
	\begin{flalign}
		H(L_{1}|Z^n,M_{1}) \leq n[R_{l1}-I(U_1;Z)]+n\delta_1\label{dd12}
	\end{flalign}
	where $\delta_1 \to 0$ as $n\to \infty$ if 
	\begin{flalign}
		R_{l1} \geq I(U_1;Z).\label{dd13}
	\end{flalign}
	
	Upon substituting (\ref{dd11}) and (\ref{dd12}) into (\ref{dd10}), we show that over all codebook generations, with probability $\geq 1-\epsilon_{0}$,  
	\begin{flalign*}
	I(M_1;Z^n)
		&\leq \epsilon_1 + n\delta_1 \\
		&\overset{\text{(c)}}{=}n\tau,&
	\end{flalign*}
	where (c) follows by defining $\tau \coloneqq \frac{\epsilon_1}{n} + \delta_1$.
	
	Similarly, for the individual secrecy of $M_2$, we show that if 
	\begin{flalign}
		R_{l2} \geq I(U_2;Z), \label{dd14}
	\end{flalign}
	then over all codebook generations, with probability $\geq 1-\epsilon_{0}$, $I(M_2;Z^n)\leq n\tau$.
		
	\textbf{Achievable rate region:} Considering only decoding error, we know that by satisfying (\ref{dd7})--(\ref{dd9}), we show the existence of at least one decoding-good codebook that satisfies (\ref{dec}) over all codebook generations. However, this argument might not be sufficient in showing the existence of codebooks that satisfy both (\ref{dec}) and (\ref{sec}), i.e., being both decoding-good and secrecy-good. From the analysis of individual secrecy in this appendix, we show that $\geq 1-\epsilon_{0}$ of the codebooks generated are secrecy-good. This incurs a problem due to the possibility of all decoding-good codebooks not being secrecy-good, indicating that a secrecy code does not exist. 
	
	We hereby prove that for all sufficiently large $n$, there exists at least one codebook that is both decoding-good and secrecy-good. Let $\mathcal{C}_n$ be the set of all codebooks for a fixed $n$. Each codebook $\mathcal{C} \in \mathcal{C}_n$ has the proability of decoding error $P_{\text{e},i}^{(n)}(\mathcal{C})$, for all $i \in {1,2}$. Given an arbitrary decoding error $\epsilon>0$, define the following:
	\begin{itemize}[leftmargin=*]
		\item Set of decoding-good codebooks $\mathcal{C}_n^1 \coloneqq \{\mathcal{C} \in \mathcal{C}_n: 0 \leq P_{\text{e},i}^{(n)}(\mathcal{C}) \leq \epsilon\}$; 
		\item Set of decoding-bad codebooks $\mathcal{C}_n^2 \coloneqq \{\mathcal{C} \in \mathcal{C}_n: \epsilon < P_{\text{e},i}^{(n)}(\mathcal{C}) \leq 1\}$; 
		\item $\alpha \coloneqq \min\limits_{\mathcal{C} \in \mathcal{C}_n^1} P_{\text{e},i}^{(n)}(\mathcal{C})$ ; and 
		\item $\beta \coloneqq \min\limits_{\mathcal{C} \in \mathcal{C}_n^2} P_{\text{e},i}^{(n)}(\mathcal{C})-\epsilon$.
	\end{itemize}
	This also gives us the fraction of decoding-good codebooks $f_d = \frac{|\mathcal{C}_n^1|}{|\mathcal{C}_n|}$. Then, the average probability of decoding error over all codebook generations is
	\begin{flalign*}
		\mathrm{P}(\mathcal{E}) &= \frac{1}{|\mathcal{C}_n|}\sum\limits_{\mathcal{C} \in \mathcal{C}_n} P_{\text{e},i}^{(n)}(\mathcal{C})&\\
		&= \frac{1}{|\mathcal{C}_n|}\sum\limits_{\mathcal{C} \in \mathcal{C}_n^1} P_{\text{e},i}^{(n)}(\mathcal{C}) + \frac{1}{|\mathcal{C}_n|}\sum\limits_{\mathcal{C} \in \mathcal{C}_n^2} P_{\text{e},i}^{(n)}(\mathcal{C})&\\
		&\geq \frac{|\mathcal{C}_n^1|}{|\mathcal{C}_n|}\alpha + \frac{|\mathcal{C}_n^2|}{|\mathcal{C}_n|}(\beta + \epsilon)&\\
		&= f_d\alpha + (1-f_d)(\beta + \epsilon)&\\
		&=\beta+\epsilon-f_d(\beta+\epsilon-\alpha).&
	\end{flalign*}
	Rearranging the terms, 
	\begin{flalign*}
		f_d &\geq \frac{\beta+\epsilon-\mathrm{P}(\mathcal{E})}{\beta+\epsilon-\alpha}&\\
		&\geq\frac{\beta+\epsilon-\mathrm{P}(\mathcal{E})}{\beta+\epsilon}&\\
		&=1-\frac{\mathrm{P}(\mathcal{E})}{\beta+\epsilon}&\\
		&\geq 1-\frac{\mathrm{P}(\mathcal{E})}{\epsilon}.&
	\end{flalign*}
	Combining with the fact that the fraction of secrecy-good codebooks $\geq 1-\epsilon_{0}$, the probability of decoding-good and secrecy-good codebooks $\mathrm{P}(\text{decoding-good}\cap\text{secrecy-good})$ is 
	\begin{flalign*}
		&\mathrm{P}(\text{decoding-good} \cap\text{secrecy-good})&\\
		&=1-\mathrm{P}(\text{decoding-bad} \cup\text{secrecy-bad})&\\
		&\geq 1-\mathrm{P}(\text{decoding-bad})-\mathrm{P}(\text{secrecy-bad})&\\
		&\geq 1-\left(\frac{\mathrm{P}(\mathcal{E})}{\epsilon}+\epsilon_{0}\right).&
	\end{flalign*}
	Thus, the number of decoding-good and secrecy-good codebooks $\mathcal{C}_{n, \text{good}}$ is 
	\begin{flalign*}
	\mathcal{C}_{n,\text{ good}}
	&=|\mathcal{C}_n|\mathrm{P}(\text{decoding-good} \cap\text{secrecy-good})&\\
	&\geq |\mathcal{C}_n|\left[1-\left(\frac{\mathrm{P}(\mathcal{E})}{\epsilon}+\epsilon_{0}\right)\right].&
	\end{flalign*}
	Notice that $\mathrm{P}(\mathcal{E}) \to 0$, $\epsilon_{0} \to 0$ and $|\mathcal{C}_n| \to \infty$ as $n \to \infty$. Hence, $\mathcal{C}_{n,\text{ good}} > 1$, showing the existence of at least one codebook that is both decoding-good and secrecy-good.
	
	The individual secrecy rate region can then be obtained from the following constraints on:  
	\begin{itemize}[leftmargin=*]
		\item the non-negativity of rates; 
		\item the rate constraints which enable Wyner secrecy coding (\ref{l3a});
		\item the achievability conditions (\ref{dd7})--(\ref{dd9}); and
		\item the individual secrecy conditions (\ref{dd13}), (\ref{dd14}). 
	\end{itemize}
	Applying the Fourier-Motzkin procedure~\cite[Appendix D]{cref6} to eliminate the terms $R_{l1}$ and $R_{l2}$ we obtain the individual secrecy rate region $\mathcal{R}$ in Theorem~\ref{theorem1}. 
\end{IEEEproof}

\bibliographystyle{IEEEtran}
\bibliography{citations}
%%
%% where we here have assume the existence of the files
%% definitions.bib and bibliofile.bib.
%% BibTeX documentation can be obtained at:
%% http://www.ctan.org/tex-archive/biblio/bibtex/contrib/doc/
%%%%%%

%% Or you use manual references (pay attention to consistency and the
%% formatting style!):
%\begin{thebibliography}{9}
	
%\bibitem{Laport:LaTeX}
%L.~Lamport,
%  \emph{\LaTeX: A Document Preparation System,} 
%  Addison-Wesley, Reading, Massachusetts, USA, 2nd~ed., 1994. 

%\bibitem{GMS:LaTeXComp}
%F.~Mittelbach, M,~Goossens, J.~Braams, D.~Carlisle, and
%C.~Rowley, \emph{The {\LaTeX} Companion,} Addison-Wesley,
%Reading, Massachusetts, USA, 2nd~ed., 2004.

%\bibitem{oetiker_latex}
%T.~Oetiker, H.~Partl, I.~Hyna, and E.~Schlegl, \emph{The Not So Short
%  Introduction to {\LaTeX2e}}, version 5.06, Jun.~20, 2016. [Online].
%  Available: \url{https://tobi.oetiker.ch/lshort/}

%\bibitem{typesetmoser}
%S.~M. Moser, \emph{How to Typeset Equations in {\LaTeX}}, version 4.6,
%  Sep. 29, 2017. [Online]. Available:
%  \url{http://moser-isi.ethz.ch/manuals.html#eqlatex}

%\bibitem{IEEE:pdfsettings}
%IEEE, \emph{Preparing Conference Content for the IEEE Xplore Digital
%  Library.} [Online]. Available:
%  \url{http://www.ieee.org/conferences_events/conferences/organizers/pubs/preparing_content.html}

%\bibitem{IEEE:AuthorToolbox}
%IEEE, \emph{Author Digital Toolbox.} [Online.] Available:
%  \url{http://www.ieee.org/publications_standards/publications/authors/authors_journals.html}

%\end{thebibliography}

\end{document}